# Status and development of the TOP-IMPLART Project

L. Picardi[1], A. Ampollini[1], P. Anello[2], M. Balduzzi[1], G. Bazzano[1], F. Borgognoni[1], E. Cisbani[2], M. D'Andrea[3], C. De Angelis[2], G. De Angelis[2], S. Della Monaca[2], G. Esposito[2], F. Ghio2, F. Giuliani[2], M. Lucentini[2], C. Marino[1], R.M. Montereali[1], P. Nenzi[1], C. Notaro[2], C. Patrono[1], M. Piccinini[1], C. Placido2, C. Ronsivalle[1], F. Santavenere[2], A. Spurio[2], L. Strigari[3], V. Surrenti[1], A. Tabocchini[2], E. Trinca[1], M. Vadrucci*[1]

[1]Italian National Agency for New Technologies, Energy and Sustainable Economic Development
ENEA, Frascati, Italy
[2]National Institute of Health, ISS, Rome, Italy
[3]National Institute of Cancer, IFO - Regina Elena, Rome, Italy

*Corresponding author: monia.vadrucci@enea.it

*Abstract*—**The TOP-IMPLART project consists of the design and implementation of a linear proton accelerator, its control and monitoring systems for the treatment of superficial and semi-deep tumors. The energy of 150 MeV (corresponding to a penetration in tissue of about 15 cm) is a milestone in design being useful for the proton therapy treatment of almost 50% of tumors based on their position and depth (including ocular melanoma, head-neck tumors, pediatric tumors, and more superficial tumors). The capability to vary the intensity on a pulse-to-pulse basis combined with an electronic feedback system allows to get the required dose uniformity (±2.5%) reducing the number of re-paintings. In this paper the state of the art and the objectives of the TOP-IMPLART project are described within the framework of the progress of Protontherapy.**

## I. Introduction

The aim of the TOP IMPLART Project is the realization of a medium-energy proton accelerator, designed as a sequence of linear accelerating structures, for proton therapy (PT) applications.

In fact, the extended name of the acronyms summarizes the peculiarities of the project from the technological point of view to perform PT: TOP "Oncology Therapy with Protons" - IMPLART, "Intensity Modulated Proton Linear Accelerator for RadioTherapy" [1].

The project is carried on by three institutions: ENEA (Italian National Agency for new technologies, energy and sustainable economic development), ISS (Italian National Institute of Health) and IFO-IRE (National Cancer Institute in Rome).

The TOP IMPLART linear accelerator (linac) is under construction and test at the ENEA Frascati Research Center; it will deliver a 150 MeV energy proton beam, penetrating up to 15 cm in tissue, and therefore suitable for clinical treatment of ocular melanoma, pediatric, head-neck and more superficial tumors.

The machine is a high frequency full linear accelerator based of a modular system composed by SCDTL (Side Coupled Drift Tube Linac) structures up to the first clinical energy (71 MeV, valid for the treatment of uveal melanoma) and CCL (Coupled Cavity Linac) modules up to 150 MeV. It is an innovative machine compared to commercial ones and this requires development of original techniques in accelerating machine, radioprotection, dosimetry, and radiobiology, which are tested during the embodiment phase. In fact, the project also includes the development of specific concepts of radio-protection, beam monitors and dosimetric systems, specific dose delivery systems (beam delivery) and the equipment related to irradiation stations for radiobiology studies on cells and animals.

At present, the accelerator is operative up to the energy of 35 MeV [2] and the sequence of accelerating modules up to 71 MeV is under construction. The prototype, originally funded by the Italian Ministry of Health, is now financially supported by the Local Government of the Regione Lazio and by the commitment of the staff and infrastructure of the ENEA Research Center in Frascati during the entire development period.

The TOP IMPLART accelerator being a modular system could be extended to higher energies (200 - 230 MeV) by adding new CCL type segment of the LINAC.

## II. State of the Art

PT belongs to the broader field of hadron radiotherapy (HRT) in which also light ions like carbon or helium are mainly used. PT is used since several years for some diseases (uveal melanoma, and tumors of the skull base and spine:

chordoma, sarcomas and meningiomas), but also for prostate, lung, liver, esophagus and head and neck-cephalic tumours. It has a recognized benefit in all cases where the disease is well localized and adjacent to critical normal organs (i.e. radiosensitive) to be saved, particularly in pediatric treatments aiming at the reduction of late effects (mainly second tumours) in normal tissues.

The rationale of the use of HRT lies in the ballistic properties and in the spatial selectivity of these particles, due respectively to the presence of the Bragg peak at the end of the particle path and to their low lateral diffusion, and especially for carbon and helium ions in their high Relative Biological Effectiveness.

*A. Current accelerators for RT applications*

The modern PT begun in 1957 at the cyclotron of the University of Uppsala (Sweden) and in 1961 at Massachusetts General Hospital (Boston), using the 160 MeV/u synchro-cyclotron of the Harvard University. Basic techniques used in HRT were developed in these facilities. In 1990 the first hospital-based PT facility was commissioned at Loma Linda University Medical Center in USA and the first hospital-based heavy-ion facility, HIMAC, was constructed at NIRS (National Institute of Radiological Sciences, Japan), which has been operated since 1994. Now several hospital-based HRT facilities are available worldwide, utilizing protons and carbon-ions. Until the year 2017 almost 200'000 patients have been treated worldwide with charged particle beams (more than 85% have been treated with protons) [3]. The HRT growth is related to the development of both the beam-delivery and accelerator technologies. All HRT facilities that have been constructed and have been in service in the world are based on circular accelerators. Generally cyclotrons are preferred for protons, while synchrotrons for ion/proton facilities. IBA, a Belgian cyclotron company is the market leader worldwide in PT cyclotrons. In Europe the GSI (Gesellschaft für Schwerionenforschung) study was used for the construction of the HIT facility at Heidelberg (Germany), based on a synchrotron provided with two horizontal fixed ports and one rotating gantry for both proton and carbon-ion RT; treatments have been successfully carried out since 2010.

In Italy the CATANA cyclotron has been active since 2002 in the research laboratory of the Italian National Institute of Nuclear Physics for the treatment of the uveal melanoma [4] and more recently other two HRT centers were activated: CNAO (in 2014, resulted by the European PIMMS study as the MedAustron center in Austria) [5] which operates with a synchrotron for protons and carbon ions, and APSS in Trento (2017) which operates with a commercial cyclotron [6].

*B. Progress beyond the state of the art for PT accelerators*

The HRT community well recognizes that progress in HRT must pursue the following developments: 1) adaptive cancer therapy, 2) treatment with hypo-fractionation/gating; 3) more compact and/or advanced machines. These developments are necessary due to the increase of precision of conventional RT, in order to decrease cost and size of the HRT facilities. Several companies have offered complete carbon-ion systems; these include the Siemens centers at Marburg and Shanghai, and several recent projects in Japan and other eastern countries involving Hitachi, Mitsubishi, Sumitomo and Toshiba [7]. The accelerator companies and research laboratories are pushed to design novel PT machines, miniaturized and less expensive but at the same time with outstanding performances in therapeutic use. This is the case of Superconducting Synchrocyclotrons (MEVION), or Cyclotrons (COMET), and of compact synchrotron (ProTom).

Proton linear accelerators (PLA) have also been proposed for the first time in 1991, by Hamm et al. [8, 9]. Proton and ion linacs are technically well assessed and sometimes are also parts of electron accelerating structures. However these accelerators were and are built mainly for very powerful applications, requiring the acceleration and handling of high current, high energy beams. Circular machines cannot accelerate particle beams with so few losses as linacs.

The low intrinsic beam losses of a linac and the possibility of varying the beam energy by controlling the RF power without using passive systems, reduce the leakage radiation and also may minimize the adverse effects plausibly correlated with secondary neutron fields. Stray radiation exposures to the patient have no known beneficial effect and are believed to increase the risk of second cancer incidence. The NCRP Report No. 170/2011 [10] describes the evolution of modern RT and how it has improved conformality reducing dose to organs near the target, but may increase stray (neutron and photon) radiation to distant organs (integral dose). Typical values of effective dose from neutron, produced out of irradiated field, range from 0.025 to 400 mSv/Gy to the tumor. Higher values are produced by the use of passive components to conform the field to the target shape, adopted in most PT centers; while lower values are measured when active scanning is performed. From the physical point of view, PLA allows the reduction of integral dose by the active spot scanning. The assessment of risk is a relevant issue for pediatric patients, who should have a long life expectancy.

The attractions of the PLA and the TOP-IMPLART program, described below, have stimulated the development of industrial initiatives currently under way. They are related to the construction of commercial proton therapy devices based on a full linac scheme [11, 12].

*C. The TOP-IMPLART project*

The TOP-IMPLART project is based on a compact proton linac dedicated to cancer therapy with the active administration of proton beams to be used directly for energy and intensity modulated (IMPT) and high repetition frequency treatment. The accelerator is designed to produce a proton beam, emitted at high repetition frequency pulses (100-200 Hz), whose characteristics (position, energy and intensity) can be varied from pulse to pulse. In this way the tumor target can be hit directing each proton pulse of the machine to a specific position of the target volume, assigning the desired dose, saving the surrounding healthy tissues to the maximum degree.

The accelerator, with energy up to 150 MeV, is under construction. Figure 1 illustrates the final configuration of the TOP-IMPLART proton therapy facility composed by the machine and three treatment rooms.

The propagation of the beam through the modules and the whole line of the machine is allowed by magnetic focusing systems (permanent quadrupole magnets or electromagnetic) of deflection (electromagnetic dipoles), correctors and on-line monitoring devices. The correct use of the beam is guaranteed by the control and radioprotection systems.

The single pulse energy is varied switching on/off the accelerating modules and varying the power in the last powered module. The output intensity can be controlled by varying the injected current as it will be described later. Other peculiarities are a small output beam size, a short beam pulse width and a high repetition rate, making it similar to the electron linacs used for cancer therapy. The TOP-IMPLART characteristics result in a modular and compact design with reduced facility and operating costs with respect to conventional accelerator employed in the cancer proton therapy systems.

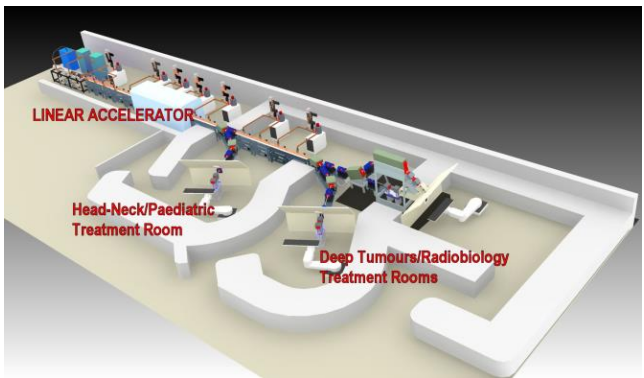

Fig.1: protontherapy facility based on the TOP-IMPLART linac.

### III. THE TOP-IMPLART ACCELERATOR

The first section of the accelerator (up to 35 MeV) has been commissioned and is currently operational producing a 50 µA current pulse at 35 MeV [13]. The second section (up to 71 MeV) is under construction and it is expected to be operational by the end of 2019.

In this paragraph the TOP-IMPLART accelerator, the main systems of the entire machine and its two proton beam extraction lines are described.

#### A. PL-7 Injector linac

The initial beam acceleration of the TOP-IMPLART accelerator is produced in a linac manufactured by the AccSys-Hitachi company, the PL-7 model, originally designed for radioisotopes production [14].

The main subsystems are a 30 keV duoplasmatron source giving a current up to 10 mA, an electrostatic einzel lens focusing the beam into the RFQ, a 2.3 meters long RFQ resonator enclosed in a vacuum chamber and equipped with a coaxial drive loop and a 1.5 meters long Drift Tube Linac (DTL) with a coaxial drive loop. The RFQ accelerates the beam of H+ ions to 3 MeV and the DTL accelerates the 3 MeV beam to 7 MeV (figure 2). The operation frequency is 425 MHz. The DTL, RFQ and injector vacuum chambers are mounted and aligned on a single rigid support structure. Two Model 12TW350 RF amplifiers connected to the RFQ and DTL by coaxial RF power feeds can supply up to 320 kW of peak RF power to each stage of the accelerator.

This accelerator was chosen due to its relatively high RF frequency and output energy and the possibility to customize it by adding to the standard high current PL7 model two elements for its use as injector for the ENEA low current proton therapy linac: an insertable beam aperture, reducing the current to the levels required by protontherapy and the possibility to vary the beam current also pulse by pulse pulsing the power supply of an einzel lens placed before the RFQ.

Table I reports the current settings of the PL-7. The output beam is injected in to a Low Energy Beam Transport line (LEBT) where two couples of quadrupoles arranged in a FODO lattice match the beam to the first booster section. A vertically bending magnet is placed between the couples to extract the injector beam for *In-vitro* radiobiology experiments. When the magnet is not engaged, the beam is injected in the booster sections for acceleration up to the energy of 150 MeV.

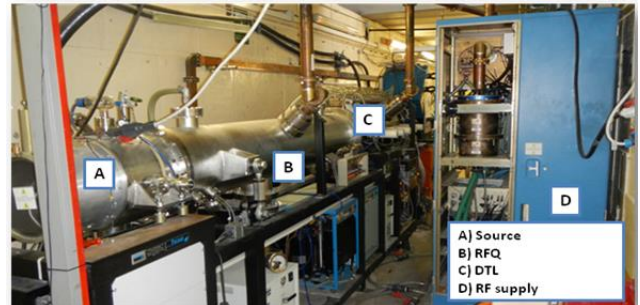

Fig. 2: the PL7 Injector. A) Source, B) the radiofrequency quadrubole, C) the drift tube linac, D) the radiofrequency supply.

TABLE I
TOP-IMPLART INJECTOR MAIN CHARACTERISTICS

| Parameter | Value | Unit |
|---|---|---|
| Ion injector output energy | 28 | keV |
| Current output (pulsed) (max) | 1.5 | mA |
| Beam pulse width | 15-80 | µs |
| Beam pulse repetition rate | 10-100 | Hz |
| Horizontal unnormalized emittance (rms) | 1.1 | π mm-mrad |
| Vertical unnormalized emittance (rms) | 1.2 | π mm-mrad |
| Energy spread (half width) | 95 | keV |

The transverse emittance is compatible with the transverse acceptance of the booster accelerator, that means that the four quadrupoles LEBT match the injector beam to the acceptance of the following accelerator and in conditions of perfect alignment the transmission is not affected by the transverse phase space of the injected beam. As to the longitudinal phase space due the large jump in RF operating frequency between the two accelerators the longitudinal acceptance area covers only the 44% of the longitudinal emittance of the PL7 beam: this means that in conditions of perfect transverse matching the transmission would be about 44% if the two machines were in adjacent positions and if the systems were matched in RF phase. A further decrease of capture down to about 10% is induced the lengthening of the beam in the LEBT. The transmission could be partly recovered inserting a buncher in the LEBT to provide the matching also in the longitudinal plane. However, we decide to tolerate this low value of transmission considering that losses occour mainly in the first section of the booster at low energies (around 7 MeV) and the current capabilities of the injector allow to reach output beam current values far higher to those required by proton therapy.

### B. High frequency booster linac

The booster linac consists of four S-band sections as shown in table III, operating at 2997.92 MHz. The first two sections employ SCDTL linac [15], because of their higher efficiency (compared to other RF linac structures) in that energy range, and consist of 8 modules. The output energy of the first two sections is 71 MeV. The last two sections employ CCL and consist of 5 modules to reach an output energy of 150 MeV.

This high frequency booster operates on RF pulses that have a width than can vary in the 1 μs - 4 μs range and a maximum repetition rate of 100 Hz.

The division in different sections reflects the RF power delivery to its modules. Each section requires between 8 and 9 MW peak power, generated by a single 10 MW peak power klystron. The RF power is divided among the modules using a power splitting and phase shifting network realized in the waveguide feed.

RF tuning of modules in a section is set at construction time and kept, during operation by a Peltier cell thermos-regulator that maintains module temperature within ±0.02 °C of the set point. An additional motorized mechanical tuner (able to accommodate ±50 kHz frequency shift) is built in each module and is controlled by an AFC (Automatic Frequency Control) loop. This tuner is used to keep each module in tune during thermal transients arising from rapid changes in the pulse repetition rate.

Figure 3 shows the machine in its current state of construction with the first 4 SCDTL modules that deliver the 35 MeV proton beam.

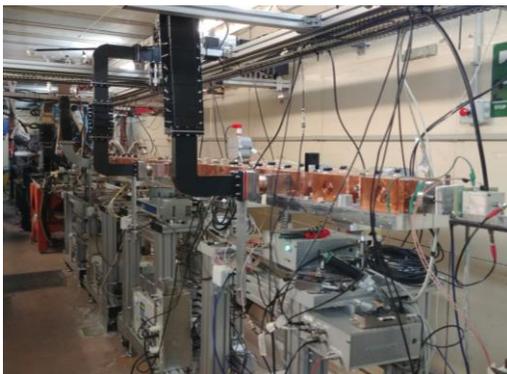

Fig. 3: View of the segment currently in operation (35 MeV) of the TOP-IMPLART accelerator.

A 4D scanning will be done by using two fast scanning magnets performing the scan in the two transversal directions, covering the depth dimension by moving the energy and intensity.

### C. Low energy vertical extraction line

The horizontal proton beam emerging from the injector (with energy in the range 3 - 7 MeV, mean current up to few pA, actively selectable) can be deflected by a dipole by 90 degree toward the zenith, in a vertical line, designed for low energy irradiation applications and specifically for radiobiology studies. In fact, it is widely recognized that the biological response is greatly determined not only by the absorbed dose, but also by the spatial and temporal distributions of the energy deposition events on a microscopic scale, i.e. by the so-called radiation quality. Better knowledge about the action of low energy protons, that are those showing the higher biological effectiveness at cellular level and present close to the end of the particle track, is therefore required. For modelling studies on the mechanisms involved in the development of carcinogenesis, irradiation of cell monolayer with proton fluences smaller than $10^6$ protons/$cm^2$, and beam uniformity on the sample around 10%, is generally required.

The vertical line, with unique instantaneous dose rate during the pulse up to hundreds Gy/s, has been recently installed downstream the bending dipole; it includes (see figure 4) a manually operable valve right after the bending dipole which allows the vacuum isolation of the line for maintenance. The valve is followed by a small tombak pipe for fine alignment of the line.

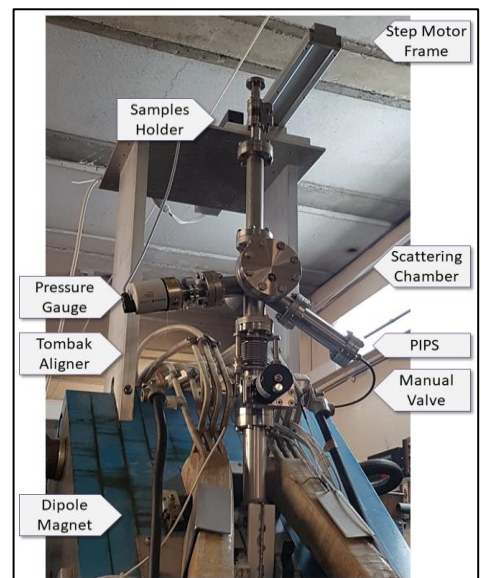

Fig.4: Vertical line for low energy irradiation; the samples holder (here for CR39 films) is in parking position.

The central element of the line is a small scattering chamber that hosts a tiny 2 um thick golden foil acting as a beam diffuser (and beam charge monitor). The upward diffused protons travel about 40 cm reaching the output mylar window, uniformly spread. The sample to be irradiated sits few mm beyond the window in a multi-sample holder; proper collimators can be inserted between line window and sample. The multi-sample holder is horizontally moved by a linear stepping motor (accuracy of 0.02 mm, up to 500 mm/s speed) and can be connected to a thermostatic system, which can maintain the samples at controlled temperature (from 4° for initial DNA damage evaluation, to 42° - 45°, for hyperthermia experiments).

The scattering chamber has two additional flanges for a pressure gauge vacuum monitor and a 150 mm$^2$ 500 µm Passivated Implanted Planar Silicon (PIPS) Partially Depleted Detector (Canberra PD150-12-500 AB) positioned at 20 cm from the golden target, and 120 degree respect to the direction of the beam.

The PIPS will be used as real-time fluence monitor: it counts the few protons per pulse scattered by the golden target at the backward angle and at the same time their energy distribution. The characterization and calibration of the line is in progress.

## IV. ACCELERATOR DIAGNOSTICS AND BEAM MONITORING

Different monitoring devices are under development for the peculiar pulsed proton beam accelerator.

For the purpose of accelerator commissioning [16] and of the proton beam characterization, set-up and operation different devices, both interceptive and non-interceptive, provide online measurements of the beam position and intensity. In this paragraph they are listed and described.

### A. Beam Monitoring for Machine Commissioning and Operation

Currently, 3 AC current transformers (ACCT), from two different manufacturers, are installed along the accelerator and operate in vacuum providing online information on the beam intensity and transmission efficiency through different sections of the machine. The first couple of ACCTs belongs to the diagnostic system of the AccSys-HITACHI injector and monitor the beam intensity at the RFQ and DTL output. The typical current in the injector is 2 mA and the detector nominal gain is $10^3$ V/A. The third ACCT is installed at the end of the LEBT for input current measurement before the SCDTL section. This detector is manufactured by Bergoz Instrumentation and provides, with calibrated amplifiers, an overall gain of $10^3$ V/A over a 1MΩ load, with a negligible droop for µs-duration pulses. With transmission efficiency in the LEBT over 90%, the current level at the SCDTL input is ≈ 1.5 mA, within the Bergoz ACCT operating range. With an axial length of 20 mm, this ACCT is a candidate for beam monitoring in the inter-tank section between SCDTL4 and SCDTL5.

Presently, a second ACCT by Bergoz Instrumentation is installed after SCDTL4 and operated in air. It is the first of a set of detectors used, during the commissioning of the accelerator for beam characterization at the exit of the last SCDTL module, namely a Faraday Cup (FC), a Cavity Intensity Monitor and a Fluorescent Screen coupled with a CCD camera (FS). With the exception of the FS, which is the only detector dedicated to profile measurement, the other devices can be used simultaneously for comparative measurement of the beam intensity and stability.

The beam current in the SCDTL section ranges between 50 µA, for commissioning purposes, and 1 µA, for radiobiology experiments and, eventually, patient treatment, close to the sensitivity and resolution limit of the ACCT. Comparison with the FC shows that the current transformer in the present configuration can be used to monitor the beam pulse by pulse intensity with accuracy comparable to the FC up to a current as low as 20 µA. Lower currents (down to ≈ 10 µA) can still be measured but the ACCT underestimates the beam stability, evaluated as σ / µ, by nearly a factor 2. At intensity typical for radiobiology experiments the current transformer can only be used to monitor the average beam current.

TABLE II
COMPARISON OF ESTIMATED BEAM STABILITY (σ / µ)

| Current level [µA] | σ / µ Faraday Cup [%] | σ / µ ACCT [%] |
|---|---|---|
| 25 | 6 | 7 |
| 10 | 6 | 12 |
| 2 | 22 | 60 |

We are therefore now testing a prototype of Cavity Intensity Monitor (CIM), a non-interceptive detector to be housed, next to Bergoz ACCT, in the inter-tank section between SCDTL4 and SCDTL5 for online monitoring of very low currents. The CIM is a passive RF cavity tuned at 2997.92 MHz equipped with a magnetic field pickup. The induced field intensity is linearly proportional to the inducing beam current; with the current prototype currents as low as 0.2 µA have been detected.

Imaging of the beam spot is carried out with an alumina fluorescent screen applied immediately after the vacuum window and a Basler ACE camera with a monochrome CCD sensor to digitally acquire the spot position and size. The camera is hardware triggered for synchronous acquisition and exposure control; the CCD sensor is 3.7 mm x 2.8 mm and has a pixel dimension of 5.6 µm x 5.6 µm; the pixel bit depth is 12 bits. The output beam is elliptical, with the horizontal axis larger than the vertical one. This is consistent with the quadrupole FODO arrangement in SCDTL structures. Typical beam size is $\sigma_x$=1.5 mm and $\sigma_y$=1.1 mm.

### B. 1D Ionization Chambers for fast beam diagnostic

Two small and thin integral 1D ionization chambers (IC_A and IC_B) have been developed for the monitor of the single pulse beam charge, down to 1 pC/pulse [17].

The two chambers performs identically but have different mechanical supports and geometries: they operates at a bias voltage of 250 V and the small (few cm$^2$ area) electrodes are made of aluminized mylar (12 µm mylar, 4 µm aluminum) and are spaced by 2 mm. Their simple front-end electronics, based on a transimpedance amplifier and integrating capacitor is derived from the 2D ionization chamber shortly described later, while the readout of the capacitor transimpedance voltage is managed by a NI-CRIO modular instrumentation, integrated in the beam control system.

Chamber IC_A sits at the exit of the beam pipe while IC_B can be placed along the beam line in air. They are currently used as, fast and sensitive, beam delivered charge monitor for development and testing purposes. The high intensity of each beam pulse combined to its short duration (few μs) and small transverse size (mm) traveling the first few tens of centimeters in air, generates a large spatial charge density in the chamber gaps, which in turn may cause noticeable volume recombination processes. These effects have been evaluated comparing the response of IC_B with the charge collected by a small Faraday Cup (intensity independent) placed right after the chamber for different pulse intensities (at fixed pulse width and frequency) and at different distances from the beam pipe exit windows. At fixed position (and constant beam transverse size) the charge collected by IC_B is linear respect to the single pulse beam intensity with a coefficient of proportionality depending on the position along the beam path in air (that corresponds to different beam transverse sizes) as reported in the plot of figure 5, where the fitting curve $\frac{Q_C}{Q} = a \cdot log\left(1 + \frac{b}{(x+c)^2}\right) \cdot (x+c)^2$ (x is the distance and a=0.0023 +/- 0.0007, b=6243+/-737 e c=24.9+/-3.0) follows from [18].

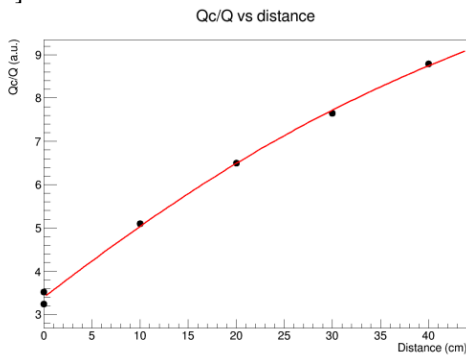

Fig. 5: Dependence of the coefficient of proportionality (Qc/Q) between the IC_B and the corresponding Faraday Cup collected charges, as a function of the distance of the IC_B from the beam exit window. The curve is a fit of the measured points; details in the text.

### C. 2D Ionization Chamber for dose delivery monitor

The intrinsically high conformation of the proton beam requires an accurate and reliable dose delivery monitor which shall measure the beam position, direction and intensity profile of each pulse. In fact, the peculiar pulsed and fast TOP-IMPLART proton beam demands for a quick and pulse-by-pulse response of the dose delivery monitor; the main requirements of this subsystem are:
- Good spatial resolution (~1/10 mm)
- Wide dynamic range ($10^4$ at least)
- Good sensitivity (~100 fC)
- Zero dead time (or near zero)
- Rapid response (< 1ms)

The real-time dose delivery monitor system is based on 2D segmented ionization chamber in Micro Pattern Gaseous Detector technology for the 2-dimensional simultaneous x/y strip readout. The chamber operates in the ionization regime to minimize saturation and discharge phenomena.

The chamber measures the single beam pulse intensity profiles simultaneously along x and y axes with spatial resolution of about 0.3 mm, sensitivity of 100 fC and a dynamic range larger than $10^4$, obtained by a dedicated electronics that automatically adapts the gain on each segment (channel) according to the amount of collected charge, which is proportional to the intensity of the beam.

Dedicated front end electronics, with multi-range logic, has been developed to optimally fit the microsecond pulsed structure of the beam [19].

A small scale prototype with sensitive area of 80x80 mm$^2$, x/y strip readout, has been characterized by calibrated dosimeters and it is routinely operated during the irradiation campaigns, as described in the following section. The 2D ionization chamber is affected by the same saturation processes described above for the 1D chambers; however at more than 100 cm from the beam pipe exit, in air, the transverse beam charge distributions (figure 6) are wide enough to make the charge density effects noticeable only at very high dose rate.

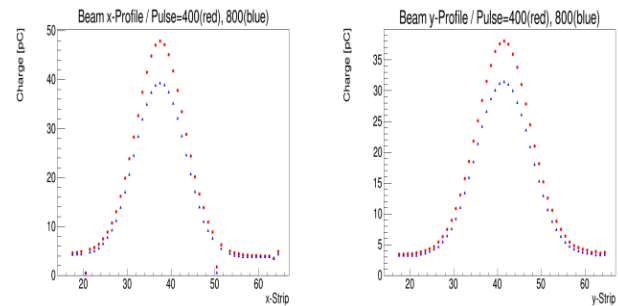

Fig. 6: x/y beam transverse profiles of single pulses measured by the 2D Ionization chamber prototype, after more than 100 cm in air.

## V. IRRADIATION CAMPAIGNS

One notable advantage of the modularity of a linac is the opportunity to exploit its beam as soon as the first acceleration structure has been installed and tuned. This way, the beam can be iteratively characterized (intensity, spatial and energy distribution, stability and reproducibility), the accelerator optimized accordingly and it can be exploited for final applications during its construction.

### A. Dosimetric characterization

Two different sessions of measurements for the dosimetric characterization in terms of the homogeneity of the laterally spread spot, the short-term stability and the dose delivery of the beam were performed; the first session was done with a beam energy of 27 MeV (24.1 MeV of maximum energy on the sample) [20]; the second one on a 35 MeV beam (31 MeV of maximum energy striking on the sample) [21]. Besides the difference in beam quality, in the period between the two measurement sessions there was an upgrade of the accelerator structure (installation of new klystron/modulator system and the implementation of resonant frequency feedback control on the accelerating modules) and a subsequent improvement of the performances were thus expected.

During the irradiations, the detectors were positioned, in free air, at a distance of ~2 m in order to obtain a lateral spread; this is required since an active scanning beam system has not been implemented yet and the beam spot diameter at the exit pipe is of only ~2 mm.

- *Beam homogeneity*

Beam homogeneity of the laterally spread spot was evaluated using EBT3 films and 2D-IC (described in paragraph IV). EBT3 Gafchromic films (Ashland ISP Advanced Materials,

NJ, USA), coming from the same batch, have an active layer of 0,028 mm placed between two 0,125 mm thick layers of polyester. Squared pieces of 2x2 cm$^2$ were placed on a custom-built dosimeter holder and positioned perpendicularly to the beam direction. The films were read using an EPSON Expression 10000XL/PRO color scanner and the beam homogeneity was evaluated as the CV% of the net pixel value (NPV) on the diameters along x and y axes; more details in[20].

- *Beam short-term stability*

For the evaluation of the *beam* short-term stability, by the estimate of its reproducibility, in terms of percent coefficient of variation, CV%, of the detector response of ten consecutive irradiations under the same experimental conditions, all the available detectors were used.

In particular, the following dosimetric systems were employed:

A high-doped p-type stereotactic field detector (Hi-pSi) DEB050 produced by Scanditronix. The silicon diode sensitive material has a thickness of 60 μm and a diameter of 0,60 mm and is encapsulated in a waterproof material. The detector was used with its axis both parallel and perpendicular to the beam axis. The diode is no longer calibrated due to its wide utilization in protons.

The microdiamond detector, mD, used was the 600019 (PTW-Freiburg, Germany), accompanied by a certificate reporting the conversion factor in 60Co by the PTW in 2017. The mD was only employed in the second measurement session, the one performed with a 35 MeV energy beam and after the improvement of the accelerating structures.

Both the silicon diode (SD) and the microdiamond detector are connected to a Keithley electrometer 6517a with no polarizing voltage.

Alanine dosimeters (Gamma Service, Leipzig, Germany), in cylindrically shaped pellets were measured with a bruker ELEXSYS EPR X-band spectrometer equipped with a high sensitivity SHQ cavity and MOSFET detectors (Best Medical, Ottawa ON, Canada)[20].

Alanine pellets and mD were also used for *dose measurements*.

Both the short-term stability and the homogeneity improved between the two measurement sessions, passing from 5% to 3.5% and from 4% to 2.6%, respectively. In figure 7 one of the first comparison of the 2D ionization chamber and alanine responses; more details are in [20, 21].

The dose assessment check was only performed on the 35 MeV energy beam, showing an accordance within 1.6% between the estimates obtained from alanine pellets and mD. This result looks acceptable, considering that it was not possible to have a perfect overlapping between the positions of the two dosimeters in the holder. Moreover, the detectors were positioned in air with their surface at the same distance from the beam pipe exit (2 m), so that their reference points were at the water-equivalent depth of 1.5 mm for alanine and at 1.0 mm for mD. This leads to a difference of 2.6% between the dose values.

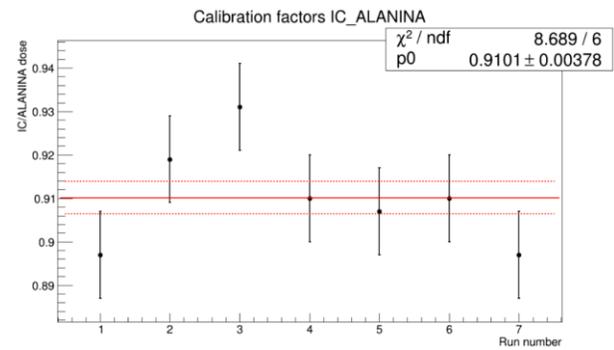

Fig.7: One of the first characterization of the 2D Ionization Chamber by mean of alanine pellets with 27 MeV proton beam; the solid line represent the mean while the dotted line correspond to the RMS from the mean.

### B. *Measurements using CR-39 track detector*

CR-39 track detectors can be used to measure the fluence of a possible low-energy proton component of the primary beam. Low energy protons interact with the front side of the CR-39 while high energy protons doesn't, starting to interact effectively with the traversed matter only after some penetration depth. As evaluated by the (SRIM [22]) code calculations, protons with energies of 12.6 MeV have a stopping range of about 1.5 mm in CR-39 (density = 1.26 g/cm$^3$). These particles, penetrating through the 1.5 mm thick CR-39, can only be stopped at the rear of the detector and do not deposit enough energy to generate tracks on the front side. Protons with energy higher than 16 MeV do not make any tracks on the CR-39.

As preliminary measures, CR-39 detectors, 18 mm (H) x 50 mm (L) x 1.5 mm (thick), were placed at the sample position and were irradiated with a proton beam of 35 MeV at the dose of 1 Gy. After irradiation, each detector was etched in 6.25 N NaOH solution at 70°C for about 4 h.

Figure 8 shows two typical fields of the etched tracks (2.31*10$^{-4}$ cm$^2$ area).

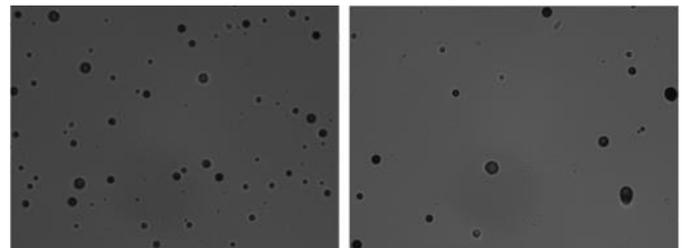

Fig. 8: Photographs (magnification 40X) of the etched tracks in a CR39 detector irradiated with 1 Gy, showing a typical field on the rear side (left panel) and front side (right panel) of the CR-39.

To evaluate the fluence of protons with energy less or equal to 16 MeV, the number of tracks included in several fields of 2.31*10$^{-4}$ cm$^2$ area can be counted for both front and rear side, using an optical microscope equipped with a CCD camera.

### C. *Novel lithium fluoride detectors for advanced diagnostics of proton beams*

Novel solid-state radiation detectors based on the photoluminescence (PL) of radiation-induced color centers (CCs) in lithium fluoride (LiF) were successfully used for advanced proton beam diagnostics [23] during the commissioning of the accelerator.

These detectors consist of LiF crystals, commercially available, and thermally-evaporated LiF films. In the crystalline lattice, the ionization induces the stable formation of the $F_2$ and $F_3^+$ electronic defects, which possess broad emission bands in the red and green spectral ranges, respectively, when optically pumped at wavelengths close to 450nm. Their PL intensity has been found to be directly proportional to the dose absorbed by the LiF material over at least three orders of magnitude [24] up to ~$10^5$ Gy. The latent PL image created by proton irradiation corresponds to the local spatial distribution of CCs which, depending on the dose amount, allows to derive the dose distribution released by the proton beam in the material. By using a conventional fluorescence microscope as reading instrument for the LiF detectors, irradiated in air with different geometries at increasing energies, useful information about both the material optical response [25] and proton beam characteristics were derived.

In the TOP-IMPLART accelerator by mounting the LiF crystals (size: 10x10x1 mm$^3$) in two different irradiation geometries, the wide PL dynamic range, combined with the high intrinsic spatial resolution, allowed obtaining two-dimensional images of both the beam transversal and longitudinal intensity distributions, each one in a single exposure. The full Bragg curves are reconstructed from the longitudinal ones, because protons release their energy and create CCs along the propagation path in number proportional to the linear energy transfer (LET). These measurements allowed quantitatively describing the proton beam characteristics in different layouts and configurations and estimating with high precision the energy of the impinging proton beams in radiobiology and dosimetric characterization experiments [20]. Figure 9 (left) shows the image of the irradiated LiF crystal obtained by the fluorescence microscope after irradiation with the proton beam in one of the experimental campaigns described in the paragraph V. Figure 9 (right) shows the intensity profile relative to the selected rectangular region (100x1600 pixel, scale 3.26 μm/pixel): the position of the Bragg peak ($X_{BP}$) retrieved from the image is at (4362±13) μm from the crystal edge ($X_0$), corresponding to a beam energy of (31.00±0.03) MeV, as derived by comparison with SRIM [22] simulations.

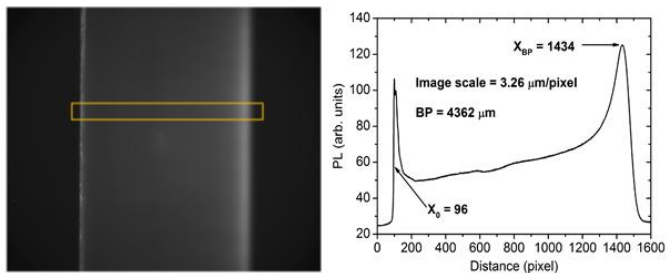

Fig. 9: (Left) Bragg curve fluorescence image stored in a LiF crystal irradiated at 1.8 m from the exit of SCDTL-4. (Right) PL intensity profile versus depth in LiF in the selected yellow marked rectangular region.

### D. Radiobiological Characterization

In order to characterize the low energy proton beam, preliminary radiobiological experiments were conducted at the TOP-IMPLART vertical transport line (before the major upgrade described above) using Chinese hamster cells, and specifically, V79 cells that are well known to be very responsive to changes in radiometric characteristics of ionizing radiation beams. For this reason, this cell line has long and widely been used at hadrontherapy facilities (e.g., at CNAO, NIRS, [26]), especially for cell killing experiments, and it is considered as a reference cell line for characterizing charged particle beams.

We performed irradiations with protons extracted in air and impinging on the cells with energy of 5 MeV (incident LET=7.7 keV/μm in MS20); the clonogenic survival was evaluated in the dose range 0.5-8 Gy. The results were found in good agreement with previous data obtained at LNL-INFN [27] using protons of comparable energy from the CN-7MV Van de Graaff accelerator.

Further irradiations to characterize the low energy proton beam were carried out on CHO-K1 cell line, evaluating clonogenic survival. The results obtained were consistent with literature data [28]. Moreover, in order to assess radiation-induced chromosome damage, an in situ protocol for micronucleus (MN) assay was developed, adapting it to the specific exposure conditions of the TOP-IMPLART vertical proton beam.

With the development of the accelerator, other radiobiological experiments have been carried out and more are planned with different proton energies to extend the radiobiological characterization of the TOP-IMPLART proton beams and to study new therapeutic approaches in Protontherapy.

In the latter direction, one of our research investigate the radiobiological optimization for combined treatment modalities. For example, the most appropriate radiobiological parameters to be applied for considering the effect of combination hyperthermia (HT) + proton therapy is ongoing also based on *In-vitro* and *In-vivo* studies. Data from clinical trials on proton therapy and hyperthermia for soft tissue sarcomas, conducted at the Paul Scherrer Institute (PSI, Villigen, Switzerland University Hospital, Zürich) might be useful for further improvement of current radiobiology studies. We have carried out *In-vitro* radiobiological experiments to evaluate HT as radiosensitizer, coupled with different qualities of ionizing radiation (X-rays and protons), on the radioresistant U-251 MG human glioblastoma cell line [29].

Irradiations at TOP-IMPLART facility were performed with 27 MeV and 35 MeV protons in the range of 1-3 Gy and the effects of radiation, alone and in combination with HT treatment, were investigated by means of clonogenic assay, cell-viability assay and MN assay (figure 10). Preliminary results, to be confirmed, suggested that HT combined with proton irradiation causes a larger radiosensitizing effect than HT combined with X-ray irradiation, as shown by clonogenic survival, by cell proliferation and by the induction of chromosome damage.

Ultimate goal is to define the optimal treatment protocol for HT + proton combination for planning the ensuing *In-vivo* investigations in a xenogenic mouse model.

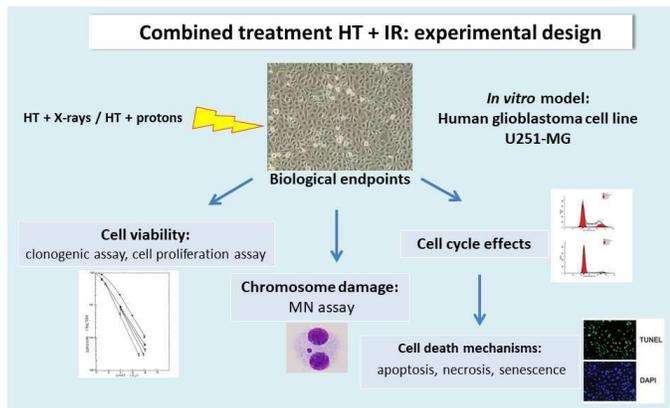

Fig. 10: Experimental design of in vitro radiobiological studies on the effects of the combination hyperthermia + ionizing radiation (IR)

## VI. DEDICATED TREATMENT PLANNING

Given the peculiarities of the TOP-IMPLART beam, we implemented an experimental tool for the calculation of its released dose. The tool is based on the convolution of a Gaussian kernel with cylindrical symmetry, obtained by Monte Carlo simulation, with an incident fluence matrix.

Monte Carlo simulation codes will be used to simulate the beam delivery line, to optimize of the resources during the design phase of the beam delivery line, to assess the delivered dose and evaluate the accuracy of calculation tool. The results of this tool will also be tested with those obtained with a commercial clinical Raystation PB Treatment Planning System for proton therapy.

The tool considers the geometry and composition of the irradiated object, the penetration of the individual beamlets, as well as the distance between the beam output and the voxel contained in the chosen calculation grid located on the target. More in details it determines the water-equivalent depths crossed by each beamlet for each element (i.e. voxel) of the calculation grid.

The tool allows performing calculation with a reduced resolution, useful when fast assessments are needed or in presence of homogeneous regions. Moreover, the dose kernels with cylindrical symmetry allow the numerical integration in all the calculation planes perpendicular to the beam axis. The spacing between these calculation planes can be modified in relation to the accuracy acceptable for dose determination. The critical zones from the accuracy point of view correspond to the Bragg peak and the distal fall, in which the dose distribution presents a very high gradient.

Our preliminary results indicate the possibility of generating analytical calculation codes to obtain sufficiently accurate dose distributions for the TOP-IMPLART experimental accelerator. For the purposes of numerical calculation, the possibility to vary the spacing between the calculation grid planes in simple geometries allows to have not only a more accurate reconstruction of the dose (<3%), but also to gain in terms of calculation time.

Dedicated treatment planning for adaptive radiotherapy are mandatory. More efforts are devoted to predic the behavior of the proton beam in the presence of geometrical / morphological variations of the target and of the organs at risk inter and intra-fraction during therapy. This issue takes advantage of studies based on CBCT and CT collected during radiotherapy treatment with photon allowing the generation of patient-specific models.

The possibility of comparing automatically generated rival plans with all the techniques currently available is also under study to identify the more appropriate and personalized treatment for each patient.